\documentclass[aps,prd,nofootinbib]{revtex4}
\usepackage{graphicx}
\usepackage{amsfonts}
\usepackage{amsmath}
\usepackage{amssymb}

\begin{document}

\title{Accretion Flow onto Ellis--Bronnikov Wormhole}
\author{R.M. Yusupova}
\email{yu.rose@mail.ru}
\affiliation{Institute of Molecule and Crystal Physics, Ufa Federal Research Centre, Russian Academy of Sciences, Prospekt Oktyabrya 151, Ufa 450075, RB, Russia}
\affiliation{Zel'dovich International Center for Astrophysics, Bashkir State Pedagogical University, 3A, October Revolution Street, Ufa 450077, RB, Russia}
\author{R.Kh. Karimov}
\email{karimov_ramis_92@mail.ru}
\affiliation{Zel'dovich International Center for Astrophysics, Bashkir State Pedagogical University, 3A, October Revolution Street, Ufa 450077, RB, Russia}
\author{R.N. Izmailov}
\email{izmailov.ramil@gmail.com}
\affiliation{Zel'dovich International Center for Astrophysics, Bashkir State Pedagogical University, 3A, October Revolution Street, Ufa 450077, RB, Russia}
\author{K.K. Nandi}
\email{izmailov.ramil@gmail.com}
\affiliation{Zel'dovich International Center for Astrophysics, Bashkir State Pedagogical University, 3A, October Revolution Street, Ufa 450077, RB, Russia}
\affiliation{High Energy Cosmic Ray Research Center, University of North Bengal,
Darjeeling 734 013, WB, India}

\date{18 June 2026}

\begin{abstract}
Study of accretion onto wormholes is rather rare compared to that onto black holes. In this paper, we consider accretion flow of cosmological dark energy modeled by barotropic fluid onto the celebrated Ellis--Bronnikov wormhole (EBWH) built by Einstein minimally coupled scalar field $\phi$, violating the null energy condition. The accreting fluid is assumed to be phantom, quintessence, dust and stiff matter. We begin by first pointing out a mathematical novelty showing how the EBWH can lead to the Schwarzschild black hole under a complex Wick rotation. Then, we analyze the profiles of fluid radial velocity, density and the rate of mass variation of the EBWH due to accretion and compare the profiles with those of the Schwarzschild black hole. We also analyze accretion to the massless EBWH that has zero ADM mass but has what we call nonzero Wheelerian mass (``mass without mass''), composed of the non-trivial scalar field, that shows gravitational effects. Our conclusion is that the mass of SBH due to phantom and non-phantom accretion increases consistently with known results, while, in contrast, the mass of EBWH decreases. Accretion to massless EBWH (i.e., to nonzero Wheelerian mass) shares the same patterns as those of the massive EBWH; hence there is no way to distinguish massive and massless cases by means of accretion flow. The contrasting mass variations due to phantom accretion could be a reflection of the distinct topology of the central objects.
\end{abstract}

%\pacs{}
\maketitle

%%%%%%%%%%%%%%%%%%  DATE  %%%%%%%%%%%%%%%%%%%

%%%%%%%%%%%%%%%%%%%%%%%%%%%%%%%%%%%%%%%%%
\section{Introduction}
\label{sec1}
%%%%%%%%%%%%%%%%%%%%%%%%%%%%%%%%%%%%%%%%%
Wormholes are topological corridors that can be used as a passageway between two distant stars or even between two distant parts of the universe. These are purely theoretical objects that are not yet ruled out by experiments; , they hence remain an intriguing curiosity among the physics community. One of the necessary conditions for the building material of wormholes is that the material should be \textit{exotic}; that is, it should violate the null energy condition (NEC) given by $\rho+p_{r}>0$, where $\rho$ is the energy density and $p_{r}$ is the radial pressure \cite{Hochberg:1998}. Classically, the NEC has so far been regarded as sacrosanct, and it formed one of the bases of objections to the existence of wormoles. However, the quantum Casimir effect produced in the laboratory tiny amounts of matter with $\rho<0$, and its Lorentz transform is just $\rho+p_{r}<0$, or NEC-violating ghost matter. It was also shown in an important paper that traversable wormholes actually require infinitesimally small amounts of exotic matter~\cite{Visser:2003}. However, for traveling to a distant star using wormholes, one would need to build a very long corridor made of a huge amount of exotic matter that is unavailable in the laboratory. On a cosmological scale, however, speculations of huge amounts of exotic matter are afloat after the remarkable discovery of accelerating cosmological expansion. It is speculated that such an expansion could be driven by NEC-violating dark energy acting against gravity, variants of which are phantom, quintessence matter or cosmological constant. Although the particle nature of dark energy has not yet been understood, it is still  possible to conceive of wormholes made of such matter. For instance, wormholes made of phantom matter \cite{Lobo:2013}, massless scalar fields, \cite{Ellis:1973,Bronnikov:1973}, etc. have been proposed, and they are just some of the many types of solutions available in the literature.

Einstein believed that not only gravity but also particles could be modeled by geometry. To that end, the concept of geometrical wormholes as a particle model was first envisaged by Einstein and Rosen in 1936 \cite{Einstein:1936}. After a gap of about fifty years, the subject received a tremendous impetus by the seminal work of Morris and Thorne in 1988 \cite{Morris:1988}. Today, a significant and useful amount of literature on wormholes of various types are available; some of them are artificially constructed, and some are exact solutions of general relativity. Belonging to the latter class is the Ellis--Bronnikov wormhole (EBWH), discovered independently by Ellis \cite{Ellis:1973} and Bronnikov \cite{Bronnikov:1973}, which are built by energy-violating Einstein minimally coupled scalar field $\phi$. Considerable work has already been done on this class, some are mentioned here though the list is by no means exhaustive \cite{Karimov:2020,Izmailov:2020,Blazquez:2018,Faraoni:2016,Nandi:2006,Nandi:2004,Nandi:1997}.

There is an important caveat relating to the stability of EBWH. As proven in several papers, EBWHs are \textit{unstable} to perturbations (see, e.g., \cite{Shinkai:2002,Gonzalez:2009}). This means that they should not be existing in the universe today---they should have died almost as soon as they were born. In that case, what is the use of studying accretion to such an object? However, some time ago, the question of (in)stability was analyzed from a completely different standpoint based not on challenging the mechanism of instability but on exploiting the fact that, in general relativity, observations are dependent on the \textit{location of the observer}. It was then shown that while some observers observe instability of EBWH, there is a nonzero probability that some other observers could observe its stability from different locations \cite{Nandi:2016}. Therefore, the present accretion scenario is relevant only to the latter types of observers.

In this paper, we  study some features of accretion flow of matter onto the EBWH spacetime. Investigation of  the accretion process started with Bondi \cite{Bondi:1952}, who formulated the problem of accretion of matter onto a compact object. Later on, Michel \cite{Michel:1972} analyzed the accretion process of a steady-state spherically symmetric flow of matter into a compact object. Consequently, several authors studied accretion process of various kinds of matter onto different types of black holes \cite{Bahamonde:2015,Debnath:2015a,Debnath:2015b,Miller:2006,MartnezPas:2014,Mach:2013,Karkowski:2006,Babichev:2004,Babichev:2013,Pepe:2012,Sharif:2012,Abbas:2019,Abbas:2020,Abbas:2021} and wormholes \cite{Gonzalez:2006a,Gonzalez:2006b,Debnath:2014,Debnath:2015c,Debnath:2020,Chattopadhyay:2015,Bandyopadhyay:2021}. We should clarify that we are considering matter flow on the background of EBWH assuming that the Einstein minimally coupled scalar field $\phi$ has entirely gone into building the background EBWH \textit{geometry} and that matter is flowing on top of that geometry. This assumption ensures that there is no interaction of the scalar field $\phi$ with accreting matter. Further, we assume that the flow is taking place only on the positive attractive side of the wormhole mouth. We  first point out how the EBWH leads to the Schwarzschild black hole under a complex Wick rotation. Then, we  analyze the radial velocity profile, the density profile and the rate of mass variation of the EBWH due to matter flow and compare them with those of the Schwarzschild black hole. As a corollary, we  also display  the accretion into massless EBWH, which is the most discussed object in the literature \cite{Khaybullina:2013,Lukmanova:2018,Tsukamoto:2016,Yoo:2013,Abe:2010,Lukmanova:2016}. We  assume units in which $8\pi G=1$, $c=1$, unless specifically restored.

%%%%%%%%%%%%%%%%%%%%%%%%%%%%%%%%%%%%%%%%%
\section{Ellis--Bronnikov Wormhole}
\label{sec2}
%%%%%%%%%%%%%%%%%%%%%%%%%%%%%%%%%%%%%%%%%
We start with the Einstein field equations that follow from the action with a minimally coupled scalar field $\phi$. The action and the resulting field equations are
\begin{equation}
S = \int d^{4}x\sqrt{-g}\left[R-\varepsilon g^{\mu\nu}\phi_{,\mu}\phi_{,\nu}\right] ,
\end{equation}%
\begin{equation}
R_{\mu\nu} = \varepsilon\phi_{,\mu}\phi_{,\nu} ,
\end{equation}%
\begin{equation}
\square\phi \equiv \phi_{;\mu}^{;\mu}=0 ,
\end{equation}
where $\varepsilon$ is a constant, $\phi_{,\mu} \equiv \partial\phi/\partial x^{\mu}$, and the semicolon denotes covariant derivative with respect to $g_{\mu\nu}$. The source scalar field $\phi$ is assumed to be a ghost field, defined by $\varepsilon = -1$, that violates all energy conditions. The EBWH solution of Equations~(2) and (3) is given in~\cite{Ellis:1973,Bronnikov:1973}. We  rewrite it in isotropic coordinates
\begin{equation}
d\tau_{\text{EBWH}}^{2} = -P(r)dt^{2} + Q(r)[dr^{2} + r^{2}(d\theta^{2} + \sin^{2}\theta d\psi^{2})],
\end{equation}%
where%
\begin{eqnarray}
P(r) &=& \exp \left[ 2\epsilon +4\gamma \tan ^{-1}(2r/m)\right] , \\
Q(r) &=& \left( 1+\frac{m^{2}}{4r^{2}}\right) ^{2}\exp \left[ 2\zeta -4\gamma\tan^{-1}(2r/m)\right] ,
\end{eqnarray}%
\begin{equation}
\phi_{\text{EBWH}}(r) = 4\delta\tan^{-1}(2r/m), \quad 2\delta^{2} = 1+\gamma^{2} ,
\end{equation}%
where $m$ and $\gamma $ are integrations constants and the coordinate patch covers only the positive mass mouth or half-patch, $0 < r < \infty$, with the ADM mass given by $M = m\gamma$. Asymptotic flatness requires that $\epsilon = -\pi\gamma$ and $\zeta = \pi\gamma$. The constraint equation $2\delta^{2} = 1+\gamma^{2}$ comes from the field equations when the solution is put into them. This form of EBWH has an istropic throat radius
\begin{equation}
r_{\text{th}} = \frac{M}{2\gamma}\left[\gamma + \sqrt{1+\gamma ^{2}}\right] .
\end{equation}

The passage from $d\tau_{\text{EBWH}}^{2}$ to the metric of Schwarzschild black hole $d\tau_{\text{SBH}}^{2}$ in isotropic coordinates is possible under a combination of inversion and complex Wick rotation as~follows
\begin{equation}
r \rightarrow -\frac{m^{2}}{4r}, \quad \gamma \rightarrow -i, \quad m \rightarrow im
\end{equation}%
such that $M = m\gamma$ and the following identity is used
\begin{equation}
\tanh^{-1}(x) \equiv \frac{1}{2}\ln\left(\frac{1+x}{1-x}\right) .
\end{equation}%

These reduce $d\tau_{\text{EBWH}}^{2}$ to
\begin{eqnarray}
d\tau _{\text{SBH}}^{2} &=& -\left(\frac{1-\frac{m}{2r}}{1+\frac{m}{2r}}\right)^{2}dt^{2} + \left(1+\frac{m}{2r}\right)^{4}\left[dr^{2}\right.  \nonumber \\
&&\left. +r^{2}\left(d\theta^{2}+\sin^{2}\theta d\varphi^{2}\right)\right] ,
\end{eqnarray}
whence the throat now reduces to the Schwarzschild horizon $r_{\text{th}} \rightarrow r_{\text{hor}} = \frac{m}{2}$. Further, the EBWH (4)--(7) was shown to be the \textit{regular} variant of the singular Class I Brans--Dicke solution in the Einstein frame \cite{Izmailov:2020,Bhadra:2001,Bhattacharya:2011}, which is just the Ellis Class I solution. These mathematical novelties were first discovered in \cite{Khaybullina:2013,Nandi:2008}, to our knowledge.

%%%%%%%%%%%%%%%%%%%%%%%%%%%%%%%%%%%%%%%%%
\section{Accretion on Massive Ellis--Bronnikov Wormhole Spacetime}
\label{sec3}
%%%%%%%%%%%%%%%%%%%%%%%%%%%%%%%%%%%%%%%%%
We  consider massive EBWH (4)--(7) and follow the procedure as in Bahamonde and Jamil \cite{Bahamonde:2015}. Assuming the accreting matter to be a perfect fluid, the energy--momentum tensor is given by the following expression
\begin{equation}
T_{\mu\nu} = (\rho+p)u_{\mu}u_{\nu} + p\,g_{\mu\nu},
\end{equation}%
where $\rho$ is the energy density, $p$ is the pressure, and $u^{\mu}$ is the four-velocity that in general is
\begin{equation}
u^{\mu} = \frac{dx^{\mu}}{d\tau} = (u^{t}, u^{r}, 0, 0),
\end{equation}%
where $\tau$ is the proper time. We assume further that $u^{\theta} = 0$ and $u^{\phi} = 0$; that is, the accretion flow is radial only. Note that all components of 4-velocity, pressure, and energy density are functions of $r$ only. Since the 4-velocity must satisfy the normalization condition $u_{\mu}u^{\mu} = -1$, we find, in the case of EBWH,
\begin{eqnarray}
u^{t}: = \frac{dt}{d\tau} &=& \sqrt{1+u^{2}\left( 1+\frac{m^{2}}{4r^{2}}\right) ^{2}\exp \left[ 2\gamma \left\{ \pi -2\tan ^{-1}\left( \frac{2r}{m}\right) \right\} \right]} \nonumber \\
&& \times\exp\left[\gamma\left\{\pi-2\tan ^{-1}\left(\frac{2r}{m}\right)\right\}\right],
\end{eqnarray}%
where for simplicity we have named $u = u^{r} = \frac{dr}{d\tau}$. Due to the presence of the square root, there are two possibilities: $u^{t} > 0$ and <0, which, respectively, imply motion forward  and backward in time. The former condition is necessary to preserve causality in the process. Moreover, to study accretion, we require $u < 0$, while for any outward flows $u > 0$. In the astrophysical context, both inward and outward flows are important. The former ones lead to growth of the wormhole, while the later ones lead to jets. Using the energy--momentum conservation law defined by $0 = T_{~~;\mu }^{\mu\nu} = \frac{1}{\sqrt{-g}}(\sqrt{-g}T^{\mu\nu})_{,\mu} + \Gamma_{~\alpha\mu}^{\nu}T^{\alpha\mu}$, we find, for EBWH,
\begin{eqnarray}
&&(\rho + p)um^{-2}r^{2}\left(1 + \frac{m^{2}}{4r^{2}}\right)^{4}\sqrt{u^{2} + \left(1 + \frac{m^{2}}{4r^{2}}\right)^{-2} \exp\left[-2\gamma\left\{\pi - 2\tan^{-1}\left(\frac{2r}{m}\right)\right\}\right]}  \nonumber \\
&&\times \exp \left[2\gamma\left\{\pi + 2\tan^{-1}\left(\frac{2r}{m}\right)\right\}\right] = A_{1},
\end{eqnarray}
\noindent where $A_{1}$ is an integration constant.

By projecting the conservation law onto the 4-velocity $u_{\mu}T_{~~;\nu}^{\mu\nu} = 0$ and contracting all indices, we can find the relativistic energy flux (or continuity) equation
\begin{equation}
u^{\mu}\rho_{,\mu} + (\rho + p)u^{\mu}\,_{;\mu} = 0.
\end{equation}%

In our case, we  assume that the pressure and energy density are related by a barotropic equation of state $p=p(\rho)$. The last equation after simplification yields
\begin{equation}
\frac{\rho^{\prime}}{\rho + p} + \frac{u^{\prime}}{u} + \frac{8(r + 2\gamma m)}{m^{2} + 4r^{2}}=0.
\end{equation}%

Here, prime denotes differentiations with respect to $r$.

Integration of Equation~(17) yields
\begin{equation}
um^{-2}r^{2}\left(1 + \frac{m^{2}}{4r^{2}}\right)^{3}\exp\left[2\gamma\left\{\pi + 2\tan^{-1}\left(\frac{2r}{m}\right)\right\} + \int{\frac{d\rho}{\rho + p(\rho)}}\right] = -A_{0},
\end{equation}%
where $A_{0}$ is an integration constant, while a negative sign is introduced on the right hand side since $u<0$ on the left hand side. Now, if we combine the above equation with (15), we~obtain
\begin{eqnarray}
&&(\rho + p)\left(1 + \frac{m^{2}}{4r^{2}}\right)\sqrt{u^{2} + \left(1 + \frac{m^{2}}{4r^{2}}\right)^{-2}\exp\left[-2\gamma\left\{\pi - 2\tan^{-1}\left(\frac{2r}{m}\right)\right\}\right]}  \nonumber \\
&&\times \exp\left[-\int{\frac{d\rho}{\rho + p(\rho)}}\right] = -\frac{A_{1}}{A_{0}} \equiv A_{3},
\end{eqnarray}%
where $A_{3}$ is a constant which depends on $A_{1}$ and $A_{0}$. Due to spherical symmetry, we take $\theta = \pi/2$ (fluid flow in the equatorial plane). Furthermore, the equation of mass flux, $0 = J_{~;\mu}^{\mu} = \frac{1}{\sqrt{-g}}\frac{d}{dr}(J^{r}\sqrt{-g})$, leads to
\begin{equation}
\rho um^{-2}r^{2}\left(1 + \frac{m^{2}}{4r^{2}}\right)^{3}\exp\left[2\gamma \left\{\pi - 2\tan^{-1}\left(\frac{2r}{m}\right)\right\}\right] = A_{2},
\end{equation}%
where $A_{2}$ is an integration constant. If we divide Equation~(15) by Equation~(20), we get another useful relation
\begin{eqnarray}
&&\frac{(\rho + p)}{\rho}\left(1 + \frac{m^{2}}{4r^{2}}\right) \sqrt{u^{2} + \left(1 + \frac{m^{2}}{4r^{2}}\right)^{-2}\exp\left[-2\gamma\left\{\pi - 2\tan^{-1}\left(\frac{2r}{m}\right)\right\}\right]}  \nonumber \\
&=& \frac{A_{1}}{A_{2}} \equiv A_{4},
\end{eqnarray}%
where $A_{4}$ is another arbitrary constant.

Taking differentials of Equations~ (20) and (21) and solving together, we obtain
\begin{eqnarray}
&& \left[V^{2}-\frac{u^{2}}{u^{2}+\left\{1+m^{2}/(4r^{2})\right\}^{-2} \exp\left[-2\gamma\left\{\pi-2\tan^{-1}\left(2r/m\right)\right\}\right]}\right]\frac{du}{u}  \notag \\
&+& \left[ -\frac{64mr^{3}\left(m+2\gamma r\right) \exp\left[-2\gamma\left\{\pi-2\tan^{-1}\left(2r/m\right)\right\}\right]}{\left(m^{2}+4r^{2}\right)^{3} \left[u^{2} + \left\{1+m^{2}/(4r^{2})\right\}^{-2} \exp\left[-2\gamma\left\{\pi-2\tan^{-1}\left(2r/m\right)\right\}\right]\right]} \right.  \notag \\
&& \left. + \frac{4m^{2}-\left(6m^{2}-8r^{2}-8\gamma mr\right)V^{2}}{r\left(m^{2}+4r^{2}\right)}\right]dr = 0,
\end{eqnarray}%
where we have introduced the variable
\begin{equation}
V^{2} \equiv \frac{d\ln(\rho +p)}{d\ln\rho} - 1.
\end{equation}

By taking the two brackets in (22) equal to zero, we can find the critical point of accretion located at $r=r_{c}$. Sonic points or critical points are the points at which the velocity of the moving fluid definitely equal to the sound speed and maximum accretion rate occurs, where the flow goes through the sonic point. Thus, at the critical point we have
\begin{eqnarray}
u_{c}^{2} &=&\frac{32\gamma mr_{c}^{5}\exp \left[ -2\gamma \left\{\pi-2\tan^{-1}\left( \frac{2r_{c}}{m}\right) \right\} \right]}{\left(4r_{c}^{2}+m^{2}\right) ^{2}\left( 4r_{c}^{2}+4\gamma mr_{c}-m^{2}\right)}, \\
V_{c}^{2} &=&\frac{2\gamma mr_{c}}{4r_{c}^{2}+6\gamma mr_{c}-m^{2}}.
\end{eqnarray}%

Here, every function is evaluated at $r=r_{c}$ and $u_{c}$ is the critical speed of the flow (velocity of the flow at the critical point).

The speed of sound is found to be
\begin{equation}
c_{s}^{2} = \left.\frac{\delta p}{\delta\rho}\right\vert_{r=r_{c}} = A_{4}\exp\left[\gamma\left\{\pi - 2\tan^{-1}\left(\frac{2r_{c}}{m}\right)\right\}\right] \sqrt{\frac{4r_{c}^{2}+4\gamma mr_{c}-m^{2}}{4r_{c}^{2}+6\gamma mr_{c}-m^{2}}}-1.
\end{equation}

The rate of change of mass $\dot{M} = -4\pi r^{2}T_{0}^{r}$ due to the accreting fluid around the gravitating object of mass $M$ ($=m\gamma $) is given by \cite{Debnath:2015a}
\begin{equation}
\dot{M}_{\text{acc}}\ =4\pi A_{0}M^{2}(\rho +p),
\end{equation}%
where the dot represents derivative with respect to time.

In general, for black holes, the mass of the central object will increase for any fluid satisfying $\rho +p>0$, which accretes on the object. On the other hand, if the fluid is a phantom dark energy $\rho +p<0$, then the mass of the central object will decrease \cite{Babichev:2004,Jamil:2009,Neves:2020}. In a vacuum, mass of the BH is a fixed constant. However, in realistic non-vacuum situations, mass cannot remain fixed: accretion leads to increase/decrease in mass, while Hawking radiation leads to a decrease in mass. To connect the BH solutions with their astrophysics, one has to take into account the time dependence of BH mass. It has to be further assumed that this time dependence does not change the global geometry and the symmetries of the spacetime; therefore, spacetime metrics remain static and spherically symmetric. If one is to consider the combined effects of accretion to and Hawking evaporation from BH simultaneously, then an additional term corresponding to Hawking radiation must be added to the right hand side of Equation (27). However, it is not our concern at the~moment.

It is possible to integrate the conservation laws and obtain analytical expressions of the physical parameters. Here, for simplicity, we study the accreting barotropic fluid with an equation of state $p = \omega\rho$. Using Equations~(15), (19), and (21), we can find directly one set of solutions (the subscript ``EB'' means the central source is EBWH):
\begin{eqnarray}
u_{\text{EB}}(r) &=& -\frac{4r^{2}\sqrt{A_{4}^{2}\exp \left[ 2\gamma \left\{\pi -2\tan ^{-1}\left( \frac{2r}{m}\right) \right\} \right] -(1+\omega )^{2}}}{\left( m^{2}+4r^{2}\right) |1+\omega|}  \nonumber \\
&&\times \exp \left[ -\gamma \left\{ \pi -2\tan ^{-1}\left( \frac{2r}{m}\right) \right\} \right] , \\
\rho _{\text{EB}}(r) &=&-\frac{16A_{2}m^{2}r^{2}|1+\omega|}{ \left(m^{2}+4r^{2}\right) ^{2}\sqrt{A_{4}^{2}\exp \left[ 2\gamma \left\{ \pi-2\tan ^{-1}\left( 2r/m\right) \right\} \right] -(1+\omega )^{2}}}  \nonumber \\
&&\times \exp \left[ -\gamma \left\{ \pi -2\tan ^{-1}\left( \frac{2r}{m}\right) \right\} \right] , \\
p _{\text{EB}}(r) &=&-\frac{16\left(A_{0}-A_{2}\right)m^{2}r^{2}|1+ \omega|}{\left( m^{2}+4r^{2}\right) ^{2}\sqrt{A_{4}^{2}\exp \left[ 2\gamma \left\{\pi -2\tan ^{-1}\left( 2r/m\right) \right\} \right] -(1+\omega )^{2}}} \nonumber \\
&&\times \exp \left[ -\gamma \left\{ \pi -2\tan ^{-1}\left( \frac{2r}{m}\right)\right\}\right] ,
\end{eqnarray}

From (27), the rate of change of the mass of the EBWH due to the accretion
process for a barotropic fluid becomes
\begin{eqnarray}
\dot{M}_{\text{EB}}(r) &=&-\frac{64\pi A_{0}A_{2}\gamma^{2}m^{4}r^{2}(1+\omega )^{2}}{\left( m^{2}+4r^{2}\right) ^{2}\sqrt{A_{4}^{2}\exp \left[ 2\gamma \left\{ \pi -2\tan ^{-1}\left( 2r/m\right)\right\} \right] -(1+\omega )^{2}}}  \nonumber \\
&&\times \exp \left[ -\gamma \left\{ \pi -2\tan ^{-1}\left( \frac{2r}{m}\right) \right\} \right] .
\end{eqnarray}

Very importantly, note that the Equations~(28)--(31) above yield Schwarzschild expressions for $\gamma = -i$, as expected (see Section~\ref{sec2}). We now fix the signs of constants based on the relevant equations, which are crucial in determining the nature of profiles. For the accretion of phantom, quintessence, dust, and stiff matter to take place, we need to have $u_{\text{EB}}(r) < 0$, $\rho _{\text{EB}}(r) > 0$, and $\rho u < 0 \Rightarrow A_{2} < 0$ from Equation (20). The sign of $A_{4}$ ($>0$) is determined by the positive real values of the radical in Equation~(28). Using the barotropic equation of state $p = \omega\rho$ and Equations~(29)--(30), we get the expression for constant $A_{0} = (1+\omega)A_{2}$, whose sign then depends on the state parameter $\omega$. Using appropriate values of these constants and respecting their signs in Equations~(28)--(31), we shall analyze the velocity profile, energy density and rate of change of mass for different values of $(1+\omega)$. The remarkable result is that, while the profiles $u_{\text{EB}}(r)$ and $\rho_{\text{EB}}(r)$ change depending on the values of $A_{2}$ and $A_{4}$, the rate of change of central mass $\dot{M}_{\text{EB}}(r)$, being proportional to $A_{2}^{2}|1+\omega|^{3}\gamma^{2}$, depends only of the sign of $|1+\omega|^{3}\gamma^{2}$ since $A_{2}^{2} > 0$.

An important issue must be noted here. (we thank an anonymous reviewer for pointing out this issue. We take this opportunity to point out that Ohgami and Sakai \cite{Ohgami:2015} have proposed a method of imaging massless EBWHs surrounded by optically thin dust. We comment that they do not consider accretion, since\ their equations yield $\overset{.}{M}=$ constant, or even $0$ (\textit{see their Equation~[3.17]}), and hence not relevant to our analysis.) At the ``phantom divide $\omega =-1$'', we note that $u_{\text{EB}}(r)$ diverges, while $\rho _{\text{EB}}(r)=0$. This divergence reflects a pathology that is known to occur at the divide; viz., the fluid perturbations become divergent, meaning that the adiabatic sound speed $c_{s}^{2}$ diverges at the crossing $\omega =-1$. This has been demonstrated by Kunz and Sapone \cite{Kunz:2006}, who studied fluid perturbations ``close'' to the phantom divide characterized by $p\approx -\rho $ (equivalently, $\omega = -1+\Delta$, where $\Delta$ is infinitesimally small) and found that the behavior of the perturbations depends crucially on the prescription for the pressure perturbation $\delta p$, which diverges at $\omega =-1$ in the dark energy rest-frame. They show that crossing the divide is possible by considering other frames where $\delta p$ can be kept finite. Another study \cite{Srivastava:2006} shows that if one considers the ``scale-factor-dependent'' equation of state $p=-\rho +f(a)$, where $f$ is a function of the scale factor $a$, then crossing the divide is also possible.

\begin{figure}
\includegraphics[width=16.2 cm]{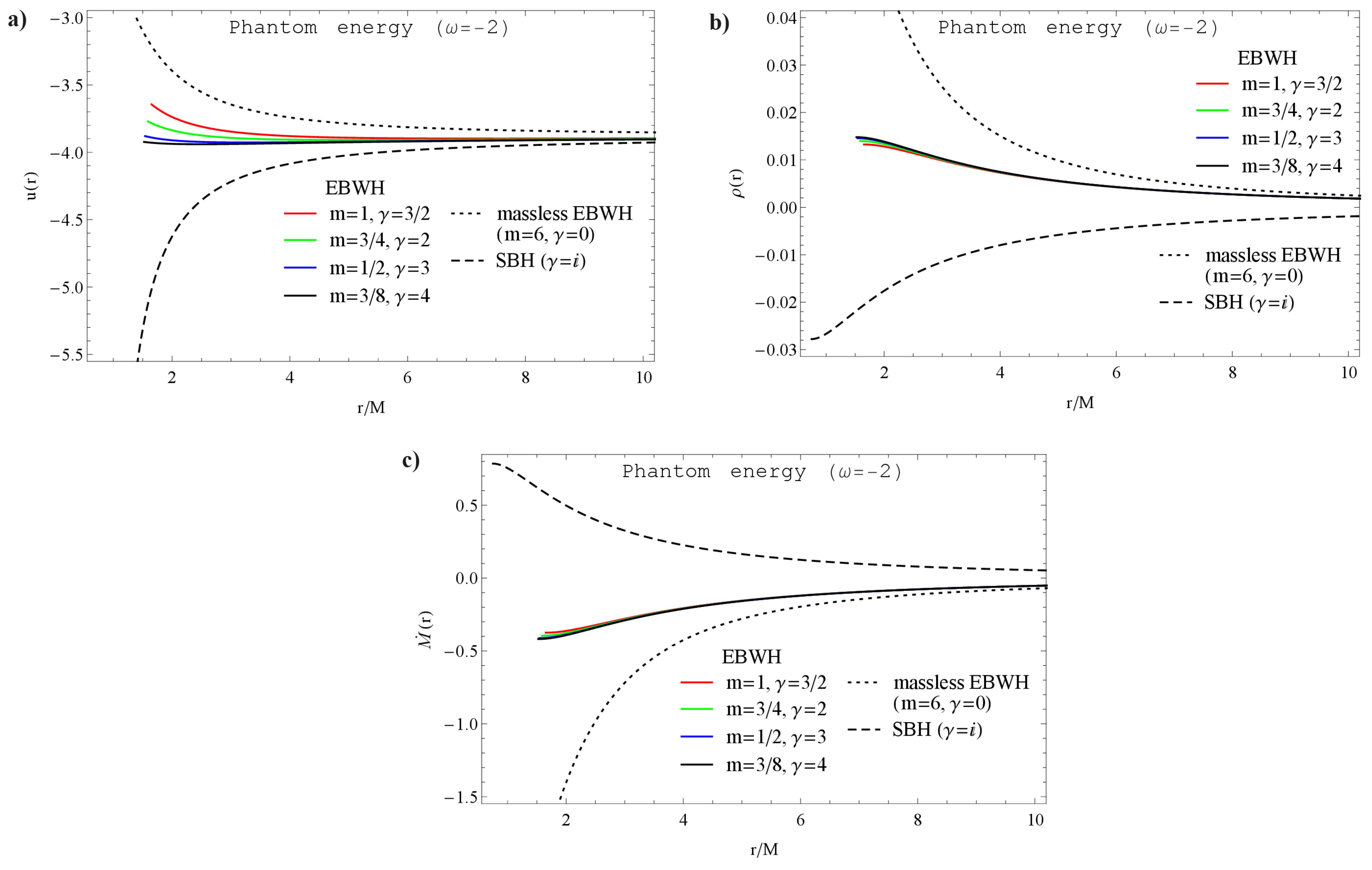}
\caption{Velocity profile (\textbf{a}), energy density (\textbf{b}) of phantom energy ($\protect\omega = -2$) and rate of change of mass (\textbf{c}) of EBWH versus $\frac{r}{M}$ for different values of $\protect\gamma $ and $m$, which satisfies $M=m\protect\gamma=3/2$. For illustration, we used the set of constants $A_{0}=1$, $A_{2}=-1$ and $A_{4}=4$.\label{fig1}}
\end{figure}

\begin{figure}
\includegraphics[width=16.2 cm]{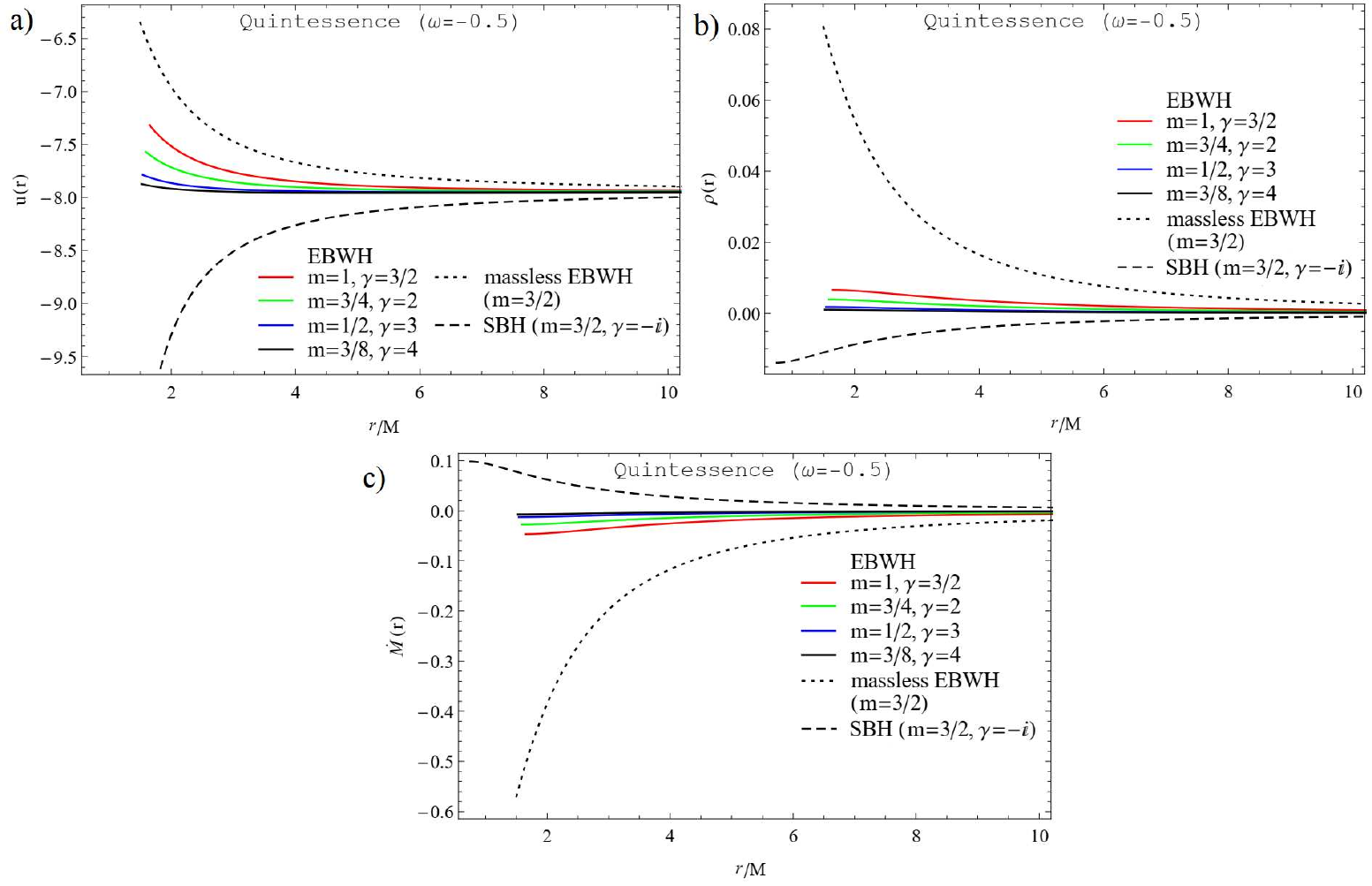}
\caption{Velocity profile (\textbf{a}), energy density (\textbf{b}) of quintessence matter ($\protect\omega =-0.5$) and rate of change of mass (\textbf{c}) of EBWH versus $\frac{r}{M}$ for different values of $\protect\gamma $ and $m$, which satisfies $M=m\protect\gamma =3/2$. For illustration, we used set of constants $A_{0}=-1/2$, $A_{2}=-1$ and $A_{4}=4$.\label{fig2}}
\end{figure}

\begin{figure}
\includegraphics[width=15.8 cm]{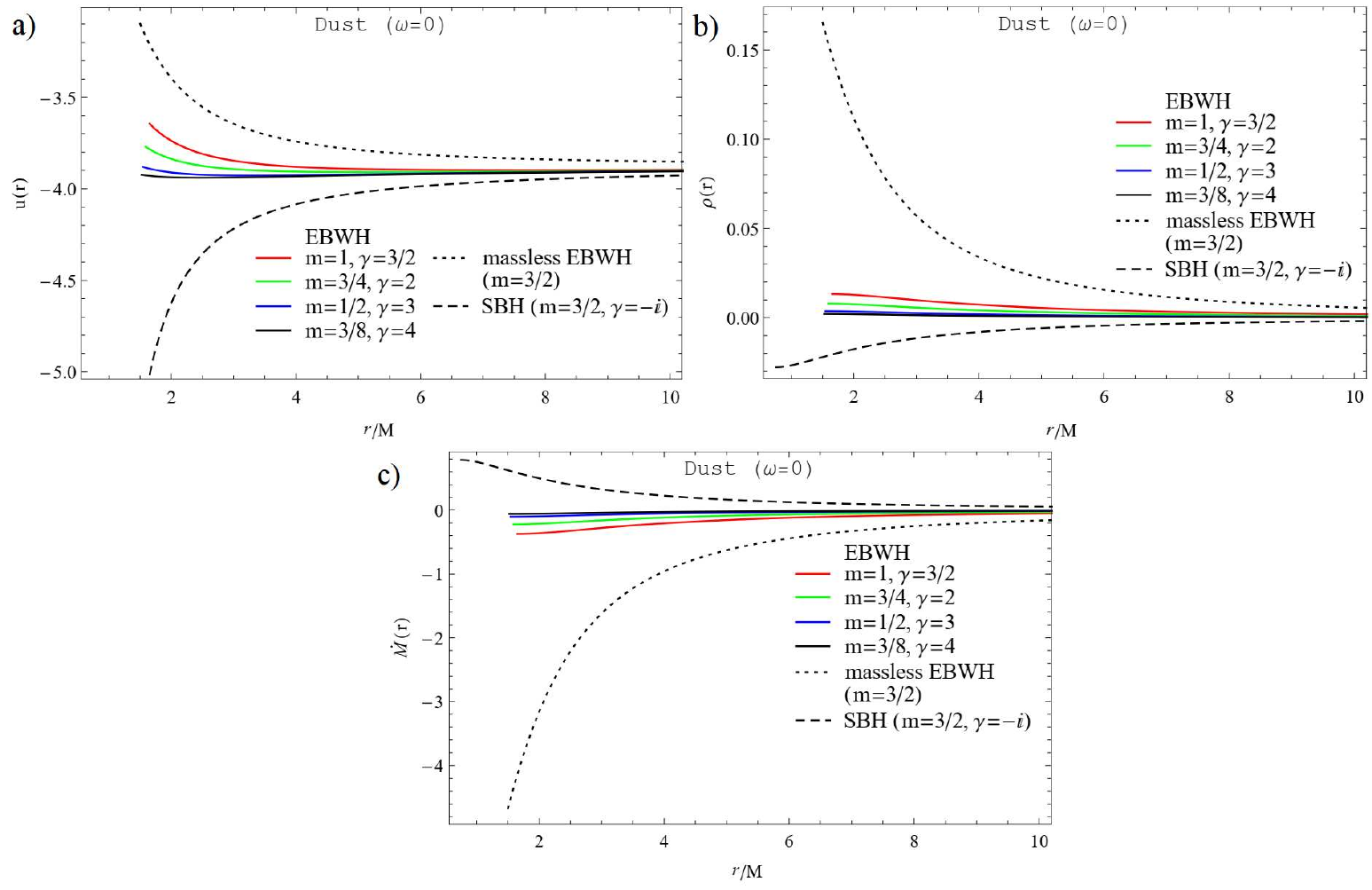}
\caption{Velocity profile (\textbf{a}), energy density (\textbf{b}) of dust ($\protect\omega =0$) and rate of change of mass (\textbf{c}) of EBWH versus $\frac{r}{M}$ for different values of $\protect\gamma $ and $m$, which satisfies $M=m\protect\gamma =3/2$. For illustration we used the set of constants $A_{0}=-1$, $A_{2}=-1$, and $A_{4}=4$.\label{fig3}}
\end{figure}

\begin{figure}
\includegraphics[width=15.8 cm]{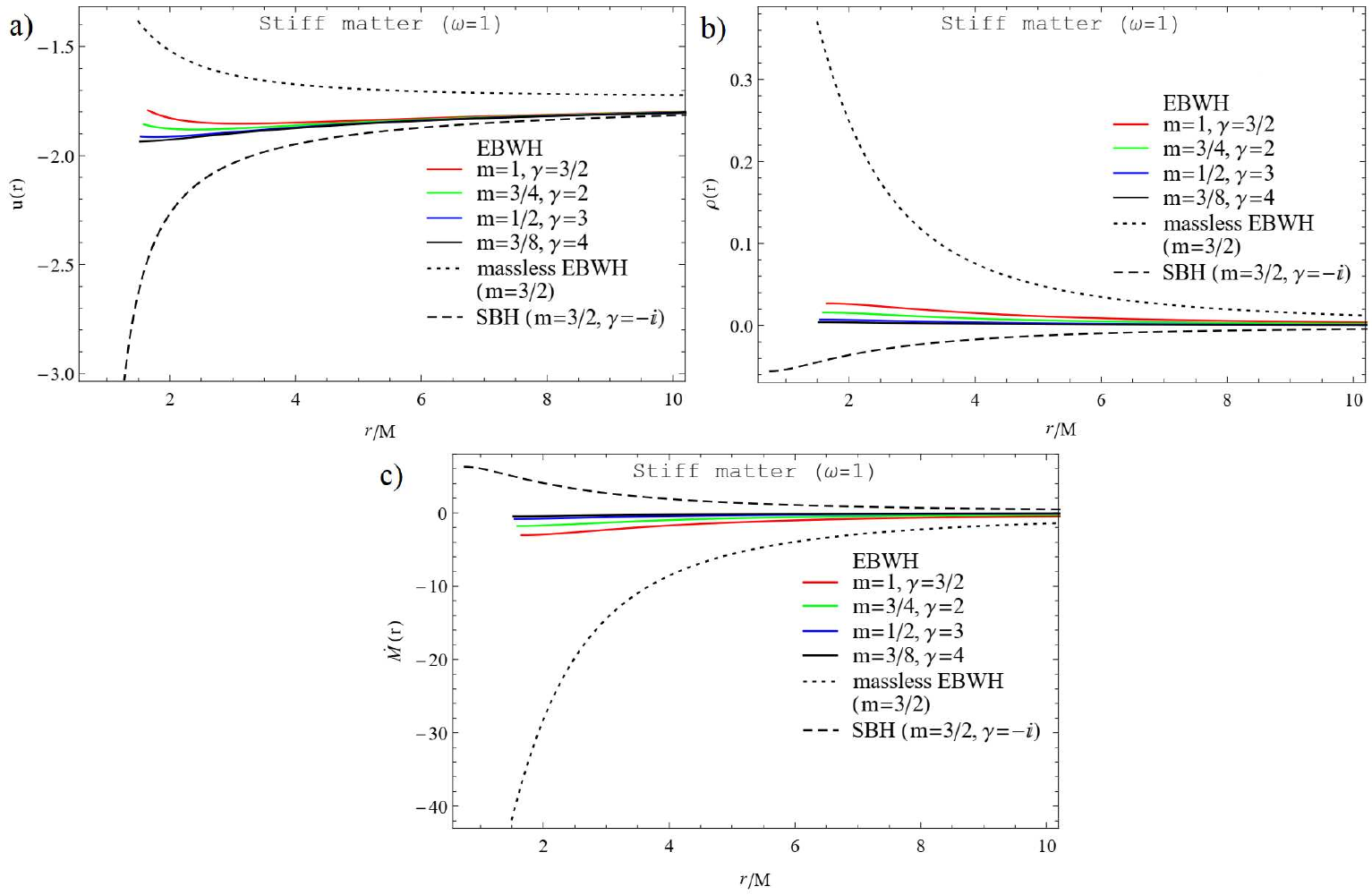}
\caption{Velocity profile (\textbf{a}), energy density (\textbf{b}) of stiff matter ($\protect\omega =1$), and rate of change of mass (\textbf{c}) of EBWH versus $\frac{r}{M}$ for different values of $\protect\gamma$ and $m$, which satisfies $M=m\protect\gamma =3/2$. For illustration, we used set of constants $A_{0}=-2$, $A_{2}=-1$ and $A_{4}=4$.\label{fig4}}
\end{figure}

%%%%%%%%%%%%%%%%%%%%%%%%%%%%%%%%%%%%%%%%%
\section{Accretion Process on Massless Ellis--Bronnikov Wormhole}
\label{sec4}
%%%%%%%%%%%%%%%%%%%%%%%%%%%%%%%%%%%%%%%%%
The EBWH massless solution follows from the massive EBWH when $\gamma = 0$ but $m \neq 0$ so that the ADM mass $M = m\gamma =0$. The metric for the massless EBWH is then
\begin{equation}
ds^{2} = -dt^{2} + \left(1 + \frac{m^{2}}{4r^{2}}\right)^{2}\left[dr^{2} + r^{2}\left(d\theta^{2} + \sin^{2}{\theta}d\varphi^{2}\right)\right] .
\end{equation}%

Under the transformation
\begin{equation}
\ell = r-\frac{m^{2}}{4r},
\end{equation}%
which maps $r\in \lbrack 0,+\infty]$ into $\ell \in (-\infty, +\infty)$, the metric goes into the popular form
\begin{eqnarray}
d\tau &=& -dt^{2} + dl^{2} + \left(\ell^{2} + m^{2}\right)\left(d\theta^{2} + \sin^{2}\theta d\varphi^{2}\right) , \\
\phi &=& \frac{1}{\sqrt{2}}\left[\frac{\pi}{2} - 2\tan^{-1}\left(\frac{\ell}{m}\right)\right] ,  \nonumber
\end{eqnarray}
where the throat occurs at  $\ell = 0$, and $m$ can be called the scalar charge proportional to the integrated total energy of the scalar field $\phi$. Each mouth reacts gravitationally; e.g., they can scatter \cite{Clement:1984}, deflect light \cite{Lukmanova:2018,Bhattacharya:2010,Izmailov:2019}, and act as gravitational lens \cite{Abe:2010,Lukmanova:2016,Gao:2019,Cheng:2021}, and hence there must be some positive energy of the scalar field on the positive side mouth causing these effects. Under a further transformation $\ell ^{2}+m^{2}=R^{2}$, the metric reduces in standard coordinates $(t,R,\theta ,\varphi )$ to the form
\begin{equation}
ds^{2} = -dt^{2} + \left(\frac{1}{1-\frac{m^{2}}{R^{2}}}\right)dR^{2} + R^{2}\left(d\theta^{2} + \sin^{2}{\theta}d\varphi^{2}\right) ,
\end{equation}%
which has the shape function $b(r)=m^{2}/R$, redshift function $\Phi = 0$, and throat at $R=m$. Note that the total integrated energy due to $\phi$ at two individual mouths is \textit{not} zero but is proportional to the non-zero scalar mass $m$ (or Wheelerian mass) on either side. This can be easily seen on the positive side by noting that the scalar field contributes a quasi-local Misner--Sharp mass, enclosed within the throat radius, which is defined by%
\begin{equation}
m(R) = \frac{R}{2}\left(1 - g^{\mu\nu}\partial_{\mu}R\partial_{\nu}R\right) ,
\end{equation}%
that yields $m(R) = \frac{R}{2}\left(1-g^{RR}\right) = \frac{m}{2}$ at $R=m$. It is now clear why the so-called massless EBWH can still react gravitationally.

It is possible to integrate the conservation laws and obtain analytical expressions for the profiles. Here, we will study again the barotropic fluid with an equation of state $p = \omega\rho$. For comparison with the massive case, it is convenient to consider the isotropic metric
\begin{equation}
ds^{2}=-dt^{2}+\left( 1+\frac{m^{2}}{4r^{2}}\right) ^{2}\left[
dr^{2}+r^{2}\left( d\theta ^{2}+\sin ^{2}{\theta }d\varphi ^{2}\right) %
\right] .
\end{equation}%

We calculate the profiles using the above massless EBWH metric ab initio regardless of its progenitor massive EBWH since now the central mass is the Wheelerian mass $m$ and there is no $\gamma$ in the metric. The results are
\begin{eqnarray}
u(r) &=& -\frac{4r^{2}\sqrt{A_{4}^{2}-(1+\omega )^{2}}}{\left(m^{2}+4r^{2}\right) (1+\omega )}, \\
\rho (r) &=& -\frac{16A_{2}m^{2}r^{2}(1+\omega )}{\left( m^{2}+4r^{2}\right)\sqrt{A_{4}^{2}-(1+\omega )^{2}}}, \\
p(r) &=& -\frac{16(A_{0}-A_{2})m^{2}r^{2}(1+\omega )}{\left(m^{2}+4r^{2}\right) \sqrt{A_{4}^{2}-(1+\omega )^{2}}},
\end{eqnarray}%
\begin{equation}
\dot{m}(r) = 4\pi A_{0}m^{2}(p+\rho )=-\frac{64\pi A_{0}A_{2}m^{4}r^{2}(1+\omega )}{\left( m^{2}+4r^{2}\right) \sqrt{A_{4}^{2}-(1+\omega )^{2}}}.
\end{equation}

%%%%%%%%%%%%%%%%%%%%%%%%%%%%%%%%%%%%%%%%%
\section{Results}
\label{sec5}
%%%%%%%%%%%%%%%%%%%%%%%%%%%%%%%%%%%%%%%%%
The following are our observations.

Figures~\ref{fig1}--\ref{fig4} represent the absolute value of the velocity profile, energy density, and rate of change of central mass for different values of the state parameter. Here, $\omega <-1$ (Figure~\ref{fig1}), $-1<\omega <-1/3$ (Figure~\ref{fig2}), $\omega = 0$ (Figure~\ref{fig3}), and $\omega = 1$ (Figure~\ref{fig4}) refer to phantom, quintessence, dust, and stiff matter, respectively.

Figure~\ref{fig1}a $\left(\omega = -2\right)$ represents radial velocity profiles of phantom fluid versus radial coordinate $r/M$ for EBWH and SBH. Values of the parameters $m$ and $\gamma $ were chosen such that $m\gamma = M$ for both the objects. We consider the central mass that is numerically the \textit{same}. For example, in the plot for massive EBWH, we take mass $M=\frac{3}{2}$ and suitable real values of $\gamma$; for massless EBWH, we take Wheelerian mass $m = \frac{3}{2}$ (no $\gamma$); for SBH, we take $m^{\prime } = \frac{3}{2}$, $\gamma = -i$. There is a steady decrease in fluid velocity showing phantom matter accretion to SBH to be consistent with the result in \cite{Babichev:2004}. On the other hand, although the profiles differ considerably in the vicinity of the sources for the same central mass $\frac{3}{2}$, the EBWH velocity profiles higher than that of SBH at all radii, showing that the phantom matter also accretes to EBWH. The same behavior also  emerges  for massless EBWH. As one moves away from the central object, all profiles tend to merge close to one another.

Figure~\ref{fig1}b represents density profiles of phantom fluid. It is seen that the density of fluid accreting to EBWH is higher than that accreting to SBH. In the case of massive EBWH, density increases in the vicinity of the throat (and horizon is case of SBH). In addition, lowering $\gamma$ also lowers the density of flow. In contrast, near the throat of massless EBWH ($r_{\text{th}} = m = \frac{3}{2}$), the density becomes maximum.

Figure~\ref{fig1}c represents the rate of change of mass $\dot{M} \sim -|1+\omega|^{3}\gamma^{2}$ against the radial coordinate $r/M$. As we can see, the accretion of phantom energy increases the mass of SBH, since $\dot{M}>0$ as a result of $|1+\omega|^{3}>0$, $\gamma^{2} = -1$, but decreases the mass of EBWH since $\dot{M}<0$ as a result of $|1+\omega|^{3}>0,\gamma ^{2}>0$. We can see from Figure~\ref{fig1}c that the property $\dot{M}<0$ is shared also by the Wheelerian mass of massless EBWH.

For quintessence (Figure~\ref{fig2}), dust (Figure~\ref{fig3}), and stiff matter (Figure~\ref{fig4}), \mbox{Figures~\ref{fig2}a, \ref{fig3}a and \ref{fig4}a} show that massless EBWH has the highest velocity of accreting fluid, and the SBH profile shows lowest velocity profiles. In the case of massive EBWH, the increase in $\gamma $ induces a decrease in velocity of the accreting fluid. Accreting matter reaches its highest values near the central source but far away from it, all the three profiles tend to bunch together.

Figures~\ref{fig2}b, \ref{fig3}b, and \ref{fig4}b represent density profiles of the accreting fluid. From the figures, it can be seen that the highest density is achieved near the central source, but again, like for the velocity, the density profiles too tend to bunch together.

%%%%%%%%%%%%%%%%%%%%%%%%%%%%%%%%%%%%%%%%%
\section{Conclusion}
\label{sec6}
%%%%%%%%%%%%%%%%%%%%%%%%%%%%%%%%%%%%%%%%%
In the above, we first pointed out a novel feature of the EBWH, namely that it reduces to the Schwarzschild black hole under the combination of a complex Wick rotation and an identity. We then analyzed the radial velocity profile, the density profile, and the rate of mass change of the massive EBWH due to the flow of dark matter with barotropic equation of state $p = \omega\rho$ and compared them with those of the SBH. As a corollary, we also discussed  the accretion onto massless EBWH, which seems to be the most discussed object in the literature (see, e.g., \cite{Lukmanova:2018,Tsukamoto:2016,Abe:2010,Ohgami:2015,Bhattacharya:2010}). The fluid flow was assumed to take place on the background of EBWH \textit{geometry} built by the massless source scalar field $\phi $ so that interaction between $\phi$ and the accreting fluid was avoided. Further, we assumed that the flow was taking place only on the positive attractive side of the wormhole mouth.

We followed the methodology of Bahamonde and Jamil \cite{Bahamonde:2015}, who studied accretion to BHs of dark matter characterized by values of the state parameter $\omega$ on either side of the phantom divide. In the present analysis, we focused on the accretion to WHs, massive and massless, which are objects topologically distinct from BHs. To our knowledge, this topic has not yet been adequately studied in the literature. While our analysis supports the known behavior of phantom accretion to BHs [20,27], the remarkable result we obtain is that the matter accretion rate $\dot{M}$ to WHs is exactly the \textit{opposite} to that of BHs: phantom and non-phantom (quintessence, dust, stiff matter) accretion increases the mass of SBH consistently with known results, while in contrast, the mass of EBWH decreases. Accretion to massless EBWH (meaning accretion to its nonzero Wheelerian mass) shares the same patterns as those of the massive EBWH; hence, there is no way to distinguish massive and massless objects by means of accretion flow. We conclude that the above contrasting behavior of accretion could be the physical signatures of the distinct topologies of the accreting central objects.

\end{document}